\documentclass[12pt]{article}

\usepackage{times}
\usepackage{algorithm,algpseudocode}
\usepackage{amsmath}
\usepackage{amsthm}
\usepackage{amssymb}
\usepackage{mathabx} 
\usepackage{graphicx}
\usepackage{color} 
\usepackage{setspace} 
\usepackage{rotating}
\usepackage{hyperref}
\usepackage{datetime}
\usepackage{natbib}
\setcitestyle{numbers,open={[},close={]}}
\usepackage[utf8]{inputenc}

\newtheorem{definition}{Definition}
\newtheorem{claim}[definition]{Claim}
\newtheorem{lemma}[definition]{Lemma}

\usepackage{todonotes}

\usepackage{multirow}
\usepackage{xspace}
\usepackage{lscape}
\usepackage{xr}

\usepackage{enumitem}

\usepackage{hyperref}
\hypersetup{
	colorlinks,
	urlcolor=blue,
	linkcolor=black,
	citecolor=black
}

\usepackage[labelfont=bf,labelsep=period,justification=raggedright]{caption}

\definecolor{gray}{rgb}{0.3, 0.3, 0.4}
\makeatletter
\renewcommand{\@biblabel}[1]{\quad#1.}

\DeclareMathOperator*{\argmin}{\arg\!\min}

\newcommand{\prefP}{{\boldsymbol{pref}(\mathcal{P})}}
\newcommand{\FSM}{FSM}
\def\NoNumber#1{{\def\alglinenumber##1{}\State #1}\addtocounter{ALG@line}{-1}}
\algnewcommand{\LeftComment}[1]{\Statex \(\triangleright\) #1}

\makeatother

\begin{document}
	
	\begin{titlepage}
		
		\title{Eliminating unwanted patterns with minimal interference}

		\author{Zehavit Leibovich$^{1,*}$, Ilan Gronau$^{1}$}
		
		\date{ }
		\maketitle
		
		\begin{footnotesize}
				$^1$Efi Arazi School of Computer Science, Herzliya Interdisciplinary Center (IDC), Herzliya 46150, Israel
				
				$^*$This research was done as part of Zehavit Leibovich's dissertation for an M.Sc degree in Computer Science. 
		\end{footnotesize}
		
		\vspace{1in}
		
		\begin{tabular}{lp{4.5in}}
			{\bf Submission type:}& Research Article
			\vspace{1ex}\\
			{\bf Keywords:}& 
			Synthetic DNA design, ~pattern matching, ~pattern elimination, ~string algorithms
			\vspace{1ex}\\
			{\bf Running Head:}&Eliminating unwanted patterns with minimal interference
			\vspace{1ex}\\ 
			{\bf Code repository:}&
			\href{https://github.com/zehavitc/EliminatingDNAPatterns.git}{https://github.com/zehavitc/EliminatingDNAPatterns.git}
			\vspace{1ex}\\ 
			{\bf Corresponding Author:}&
			\begin{minipage}[t]{4in}
				Ilan Gronau\\
				Efi Arazi School of Computer Science\\
				The Herzliya Interdisciplinary Center\\
				P.O.B. 167 Herzliya, 46150 Israel\\
				Phone: +972-9-952-7907\\
				Fax: +972-9-956-8604\\
				Email: ilan.gronau@idc.ac.il
			\end{minipage}
		\end{tabular}
		
		\thispagestyle{empty}
	\end{titlepage}
	
	\doublespacing
	\newpage
	
	\section*{Abstract}
	Artificial synthesis of DNA molecules is an essential part of the study of biological mechanisms. The design of a synthetic DNA molecule usually involves many objectives. One of the important objectives is to eliminate short sequence patterns that correspond to binding sites of restriction enzymes or transcription factors. While many design tools address this problem, no adequate formal solution exists for the pattern elimination problem. In this work, we present a formal description of the elimination problem and suggest efficient algorithms that eliminate unwanted patterns and allow optimization of other objectives with minimal interference to the desired DNA functionality. Our approach is flexible, efficient, and straightforward, and therefore can be easily incorporated in existing DNA design tools, making them considerably more powerful.

	
	

	\section{Introduction}\label{Intro}
	Synthetic biology is an emerging domain that uses engineering principles to study biological mechanisms by examining perturbations of these mechanism. This field has seen rapid growth in research and innovation in recent years \cite{IntroTrackingTheEmergenceOfBioSyn}. Many applications of synthetic biology involve artificial synthesis of DNA molecules based on some specification \cite{SynBioPutSynInBio}. An example of such an application is the pilot project announced by an initiative called the Human Genome Project-write (HGP-write) to create a virus-resistant cell by removing DNA sequences from the human genome that viruses use to hijack and replicate \cite{VirusResistentCell}. Another application is to conduct experiments to test theories, such as the experiment that confirmed that CRISPR (clusters of regularly interspaced short palindromic repeats) is used by bacteria to recognize viruses and handle future attacks. This finding later led to using CRISPR to alter the DNA of human cells like an exact and easy-to-use pair of scissors \cite{CRISPRAsDnaEdit}. These examples demonstrate that with the rapid progress in relevant technologies, it is expected that synthetic biology will be able to help resolve many key open questions in molecular biology.
	
	 In many applications, like the ones presented above, the synthesized DNA molecule is a molecule that was artificially designed to meet some requirements. The design of protein-coding sequences usually involves meeting objectives such as optimizing codon usage, restriction site incorporation, and motif avoidance. Whereas meeting only one objective can be relatively simple, meeting multiple objectives at once is a much more complicated task, and therefore, many tools heavily rely on heuristics based on random sampling \cite{DnaSynToolsReview}. One particularly challenging task in DNA sequence design is avoiding certain short sequence patterns that correspond to potential binding sites of proteins such as restriction enzymes or transcription factors. Cleaning the synthesized sequence from potential binding sites is essential when one wishes to control the function of that sequence in a cellular environment. Compared to other design objectives that try to optimize some properties, this problem involves a strict restriction: we must remove all unwanted patterns because even one occurrence of a binding site can affect the DNA function. This strict restriction, along with positive specification that one wishes to optimize, introduces a significant computational challenge.
	
	In this work, we examine the problem of eliminating unwanted sequences from a given target sequence with minimal disturbances. We start by examining the simple question of cleaning a single unwanted pattern from a target DNA sequence. We show that various versions of this problem can be solved by reduction to the well-known hitting set problem. Later, we present a dynamic programming scheme that solves a more general version of this problem that, among other things, cleans multiple unwanted patterns. All of the algorithms we present in this work are linear in the size of the input. We also provide related software tools in a public repository:\\ \href{https://github.com/zehavitc/EliminatingDNAPatterns.git}{https://github.com/zehavitc/EliminatingDNAPatterns.git}. 

	\section{Related works}\label{RelatedWork}

	\subsection{Design tools}\label{DesignTools}

	Modern DNA design tools aim to meet multiple design preferences and objectives, as reviewed in \cite{DnaSynToolsReview}. Table \ref{tab:DNASynToolComparison} summarizes the objectives that the different tools claim to achieve. All tools consider codon usage, meaning that they attempt to choose a codon for each protein amino acid based on usage statistics in the organism whose cells are used in the experiment. Considering codon usage is clearly central in experiments that involve synthetic DNA. Computationally, it is relatively simple to address using the organisms codon usage distribution. Other than codon usage, tools differ in the set of objectives they claim to address. Most tools claim to address some version of pattern elimination, either through a user-defined set of patterns or by eliminating a pre-defined set of patterns (hidden stop codons, binding sites of certain restriction enzymes, etc.).

\begin{table}[t]
	\caption{\textbf{Design features supported by different design tools.} \textmd{The features are ordered from left to right, first the codon usage optimization feature that is supported by all of the tools, then five features related to pattern elimination, then six features ordered by the number of tools supporting them. This table is adapted from Tables $1$ and $2$ from \cite{DnaSynToolsReview}}}
	\label{tab:DNASynToolComparison}
	\resizebox{\textwidth}{!}{%
		\begin{tabular}{lccllllccccccc}
			\hline
			\\
			\textbf{\begin{tabular}[c]{@{}l@{}}Gene design\\  tool\end{tabular}} &
			\multicolumn{1}{l}{\textbf{\begin{tabular}[c]{@{}l@{}}Codon \\ usage\end{tabular}}} &
			\multicolumn{1}{l}{\textbf{\begin{tabular}[c]{@{}l@{}}User-defined\\ restriction \\ site \\ elimination\end{tabular}}} &
			\textbf{\begin{tabular}[c]{@{}l@{}}Pre-defined\\ sites\\ elimination\end{tabular}} &
			\textbf{\begin{tabular}[c]{@{}l@{}}Hidden \\ stop \\ codons\end{tabular}} &
			\textbf{\begin{tabular}[c]{@{}l@{}}Motif \\ avoidance\end{tabular}} &
			\textbf{\begin{tabular}[c]{@{}l@{}}Repetitious \\ base\\  removal\end{tabular}} &
			\multicolumn{1}{l}{\textbf{\begin{tabular}[c]{@{}l@{}}GC \\ content\end{tabular}}} &
			\multicolumn{1}{l}{\textbf{\begin{tabular}[c]{@{}l@{}}Oligo\\  generation\end{tabular}}} &
			\multicolumn{1}{l}{\textbf{\begin{tabular}[c]{@{}l@{}}mRNA \\ secondary \\ structure\end{tabular}}} &
			\multicolumn{1}{l}{\textbf{\begin{tabular}[c]{@{}l@{}}Codon\\  context\end{tabular}}} &
			\multicolumn{1}{l}{\textbf{\begin{tabular}[c]{@{}l@{}}Codon\\  auto-\\ correlation \\ adjustement\end{tabular}}} &
			\multicolumn{1}{l}{\textbf{\begin{tabular}[c]{@{}l@{}}Hydropathy \\ index \\ optiomization\end{tabular}}} &
			\textbf{Reference} \\ 
			\hline
			\\
			\textbf{DNAWorks}                                                          & X & X &   &   &   &   &   & X &   &   &   &   & \cite{DNAWorks} \\
			\textbf{Jcat}                                                              & X &   & X &   &   &   &   &   &   &   &   &   &  \cite{Jcat}\\
			\textbf{\begin{tabular}[c]{@{}l@{}}Synhetic gene \\ designer\end{tabular}} & X &   & X &   &   & X &   & X &   &   &   &   &  \cite{SynGeneDesigner} \\
			\textbf{GeneDesign}                                                        & X & X &   &   &   &   &   & X &   &   &   &   & \cite{GeneDesign} \\
			\textbf{Gene Designer 2.0}                                                 & X & X &   &   &   &   &   &   &   &   &   &   & \cite{GeneDesigner} \\
			\textbf{OPTIMIZER}                                                         & X & X &   &   & X &   & X & X &   &   &   &   & \cite{Optimizer} \\
			\textbf{\begin{tabular}[c]{@{}l@{}}Visual gene \\ developer\end{tabular}}  & X & X &   & X & X & X & X &   & X &   &   &   &  \cite{VisualGeneDeveloper} \\
			\textbf{Eugene}                                                            & X & X &   & X &   & X & X &   & X & X & X &   &  \cite{Eugene} \\
			\textbf{COOL}                                                              & X &   & X & X & X & X & X &   &   & X &   &   & \cite{COOL} \\
			\textbf{D-tailor}                                                          & X & X &   &   &   &   & X &   & X &   &   & X & \cite{DTailor} \\
			\hline
		\end{tabular}%
	}
\end{table} 
	
	Gould and colleagues in \cite{DnaSynToolsReview} sought out to examine how well different tools deal with the pattern elimination objective together with other competing objectives. They took a target sequence and specified two restriction sites to be removed. They also restricted the codons that can be used such that no valid sequence of codons will eliminate the restriction sites. Thus, the design requirements cannot be met in this case. The purpose of this experiment was to see how tools behaved when posed with a pattern elimination objective that conflicts with another design requirement. Four tools (Gene Designer 2.0 \cite{GeneDesigner}, Jcat \cite{Jcat}, Eugene \cite{Eugene}, and D-Tailor \cite{DTailor}) were not tested because they do not have the option to configure this specific design objective. One tool became unresponsive (Synthetic gene designer \cite{SynGeneDesigner}), possibly because there is no feasible solution. Two tools (DNAWorks \cite{DNAWorks} and Visual gene developer \cite{VisualGeneDeveloper}) left the restriction sites. It is unclear whether the tools indicated that they could not remove the restriction sites. The remaining three tools (GeneDesign \cite{GeneDesign}, OPTIMIZER \cite{Optimizer}, COOL \cite{COOL}) removed the restriction sites using restricted codons for two amino acid. 
	
	 It seems that the tools do not expect a set of constraints that cannot be met. One of the reasons for the difficulty that existing tools have in addressing complex, and possibly conflicting, constraints is likely due to the general technique they all use. As far as we can tell, all programs eliminate unwanted patterns by scanning the DNA sequence, and each time they encounter an unwanted pattern, they choose a random  substitution (as done in \cite{GeneDesigner, OnlineToolDesignBindingSiteFreeDna}). This strategy is simple and can be effective in many cases, but it ignores the possible complexities of the pattern elimination problem. One potential problem that this approach ignores is that removing one unwanted pattern can create a new unwanted pattern. Therefore, random sampling cannot guarantee a feasible and optimal solution and might be ineffective. This becomes more problematic the more patterns you wish to eliminate. 
	 Another clear problem with how these tools address the pattern elimination problem is that they do not clearly specify the algorithm or heuristic protocol they use. Consider, for example, two of the tools that removed the restriction sites in the test described above. The article that published OPTIMIZER (\cite{Optimizer}) does not mention the algorithm used at all, and the article that published GeneDesign (\cite{GeneDesign}) only mentions that it uses a random selection of codons. 

	\subsection{Theoretical analysis of related problems}\label{TheoreticalWork}
	The patterns elimination problem first requires finding all pattern matches. There are two ways to address this problem. One is inspired by the Knuth-Morris-Pratt (KMP) \cite{KMP} algorithm, and the other is using a suffix tree. 
	The KMP algorithm finds all matches of a single pattern in a given sequence using a protocol it
	constructs based on the given pattern. The KMP protocol can be described using a simple finite state machine (FSM) that traces any given sequence and keeps in every state the longest prefix of the pattern that is also a suffix of the	sequence traced thus far. When the FSM reaches the state corresponding to the complete pattern, this indicates that a match has been found. In \cite{KMPBasedPatternMatching}, Aho and Corasick describe an efficient method for creating a FSM that is inspired by the KMP FSM and matches multiple patterns in a given sequence. The FSM they describe keeps in every state the longest prefix of \emph{one of the patterns} that is also a suffix of the sequence traced thus far. Finding all pattern occurrences using this FSM is linear in the sequence length, and it does not depend on the length or the number of patterns. Building this FSM requires a pre-processing time that is linear in the sum of lengths of all patterns. Another approach for solving the pattern matching problem is using a suffix tree \cite{SuffixTree}, which is a data structure whose nodes correspond to substrings of a given sequence and whose leaves hold indices in it. Each path in the tree from the root to a leaf corresponds to a suffix of the sequence: the leaf holds the starting position of the suffix, and the concatenation of all the nodes' substrings in the path gives the sequence of the suffix. After building the suffix tree of the sequence, all pattern matches can be found in time that is linear in the sum of lengths of all patterns by simply searching for a pattern starting at the root, as each substring is a prefix of a suffix of the sequence. 
	  
	  There have been several studies that examine theoretical and algorithmic aspects of the pattern elimination problem. Some problems have been studied and were shown to be NP-complete.   	  
	  For example, in \cite{BetterPhages} Skiena addressed the problem of minimizing the number of restriction sites while keeping the set of given genes unchanged (codon substitution is permitted only if the resulting amino acid is the same). He suggests a dynamic programming algorithm that is exponential in the length of the longest restriction site and proves that the problem is NP-complete for non-fixed restriction site lengths. Another related problem is the Unique Restriction Site Placement Problem  (URSPP) presented in \cite{RestrictionSitesForSynGen}. The objective in this problem is to allow only one restriction site for any given restriction enzyme, keep the translated sequence of amino acids unchanged, and minimize the maximum gap between adjacent restriction sites. They show that this problem is NP-complete and then suggest a heuristic algorithm that starts with eliminating all but one binding site for each restriction enzyme. They do not provide a detailed description of their algorithm and specifically how they avoid creating new restriction sites. 
	  Both \cite{BetterPhages} and \cite{RestrictionSitesForSynGen} give higher priority to avoiding changes in the translated amino acid sequence over the number or placement of restriction sites. 
	  
	  A recent study \cite{deletingPatternMultiD} addressed the problem of eliminating a single unwanted pattern in the context of $2D$ images (and multi-dimensional arrays). The results of \cite{deletingPatternMultiD} focus on the problem of deciding if a multi-dimensional array is clean of an unwanted pattern and measuring its distance from being clean. One of their results suggested a simple and efficient algorithm for eliminating a single pattern from a sequence over a binary alphabet. Our work uses the results of \cite{deletingPatternMultiD} in the one-dimensional case as a starting point for dealing with the pattern elimination problem. In Section \ref{ConnectionBetweenElimAndHitting} we extend a lemma that was proved by \cite{deletingPatternMultiD} (Lemma $18$) to establish the connection between the pattern elimination problem and the hitting set problem over the DNA alphabet. 
	\section{Definition of objectives and notations}
	 We consider a long target sequence $S$ of length $n$ over an alphabet $\Sigma$. The sequence $S$ represents the optimal version of the synthesized sequence without considering possible existence of unwanted patterns. If we wish to synthesize multiple sequences, we concatenate them into one long target sequence $S$, using a unique character to separate between individual sequences.
	 Our main objective is to clean the target sequence $S$ from occurrences of short patterns specified in the set $\mathcal{P}$. Typically, the sequences in $\mathcal{P}$ are much shorter than the target sequence $S$.	 
	 
	 We use a $1$-based indexing scheme and denote by $S_i$ the $i^{th}$ character in $S$, and by $S_{i...j}$ the substring of $S$ that begins in index $i$ and ends in index $j$. Our objective is defined by the following concepts:
	 \begin{definition}
	 	Given a sequence $S$ and a short pattern $P$ of length $k$, a $P$-match in $S$ is a substring of $S$ that is identical to $P$: $S_{i...i+k-1} = P$.
	 \end{definition}
	 \begin{definition}
	 	Given a collection of short sequence patterns, $\mathcal{P} \subseteq \Sigma^k$, a sequence $S$ is said to be $\mathcal{P}$-clean iff $S$ does not contain a $P$-match for every $P \in \mathcal{P}$.	
	 \end{definition} 
	 \begin{definition}
	 	Given a target sequence $S$ and a collection of short sequences $\mathcal{P}$, an eliminating set for $\mathcal{P}$ in $S$ is a set $E \subseteq \{1..n\}\times\Sigma$ such that substituting $S_i$ with character $\sigma$ for all pairs $(i,\sigma)\in E$ results in a sequence $S'$, which is $\mathcal{P}$-clean.
	 \end{definition}  
	 In the following sections, we describe a series of algorithms that find an {\em optimal} eliminating set under different scenarios. In Section \ref{ConnectionBetweenElimAndHitting}, we start with the simple scenario where $\mathcal{P}$ contains a single pattern $P$, and we wish to find the smallest eliminating set. In Section \ref{PositionSpecificRestrictions}, we expand the optimization criterion to consider positional-preferences for substitutions. In both sections, we consider elimination of a single pattern and thus equate the set $\mathcal{P}$ with the single pattern $P$ it contains. Finally, in section \ref{FSMIntro} we expand the discussion to the multi-pattern case and to more general optimization criteria. 
	\section{The connection between eliminating sets and hitting sets}\label{ConnectionBetweenElimAndHitting}
	We start by considering the simple problem of finding the smallest eliminating set for a given target sequence, $S$, and a single pattern, $P$. Clearly, the set of positions of any elimination set has to cover all $P$-matches. However, a set that covers all of the $P$-matches is not necessarily an eliminating set, because substituting $S_i$ may create new $P$-matches.
	Consider the following example over the binary alphabet:
	\begin{figure}[H]
		\centering
		\includegraphics[width=0.5\linewidth]{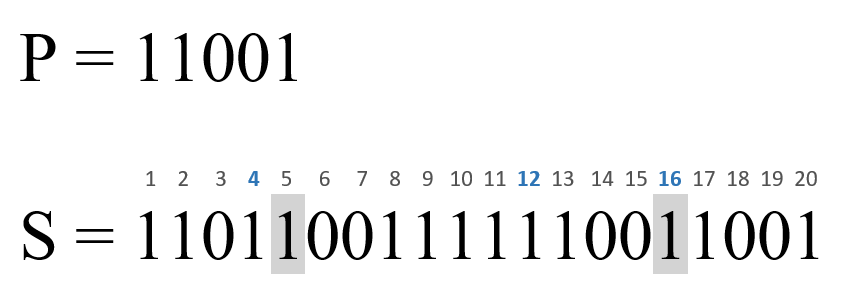}
		\caption{Eliminating pattern example}
		\label{fig:EliminatePatternExample}
	\end{figure}
	There are three $P$-matches in $S$ starting in positions $4, 12$, and $16$. If the bit in position $4$ is flipped, then the first $P$-match is eliminated, but a new one is created (starting in position $1$). On the other hand, flipping each of the bits in positions $5-8$ eliminates this $P$-match without creating a new $P$-match. The second $P$-match can be eliminated by flipping each of the bits in positions $12-16$, but flipping the bit in position $16$ also eliminates the third $P$-match, so it is clearly preferable. 
	This example demonstrates that some substitutions may eliminate an existing $P$-match but may also create a new one. The example also demonstrates that we should aim to utilize overlaps between $P$-matches in order to minimize the number of substitution. Optimal utilization of overlaps can be achieved by finding a \emph{minimal hitting set} for the set of $P$-matches.
	\begin{definition}
	Let $I=\{[l_1, r_1],...,[l_n, r_n]\}$ be a set of intervals of a sequence $S$. Let $H$ be a subset of positions in $S$. $H$ is a hitting set of $I$ if each interval $[l,r] \in I$ contains at least one position in $H$. 
	\end{definition} 	
	The minimal hitting set problem is a specific instance of the more general set cover problem, which is known to be NP-hard. However, when the sets correspond to contiguous intervals of natural numbers, this problem has a simple linear-time algorithm, which we describe in Section \ref{HittingSetAlg}. The following lemma provides a key observation to our analysis, establishing an important connection between hitting sets and eliminating sets. 
 	\begin{lemma}\label{EliminatingDnaClaim}
		If a position $j$ in $S$ belongs to a $P$-match, then substituting $S_j$ with any character can create at most one new $P$-match.
	\end{lemma}
	A version of this lemma restricted to binary sequences was proven in \cite{deletingPatternMultiD} (Lemma $18$). For completeness, we provide a detailed proof of Lemma \ref{EliminatingDnaClaim} in Section \ref{ProofOfEliminatingDnaClaim}. One important implication of this lemma is that for non-binary alphabets, the eliminating set problem is reduced to the hitting set problem, such that any hitting set can be extended to an eliminating set using the same positions. 
 	\begin{claim}\label{corollaryHittingEqualEliminating}
		If the alphabet $\Sigma$ has more than two characters, then the elimination problem of a single pattern reduces to the hitting set problem.
	\end{claim}
	\begin{proof}
		Let $\Sigma={\sigma_1,...\sigma_t}$, where $t>2$, and let $H$ be a hitting set of all $P$-matches in $S$.
		Consider an arbitrary position in the hitting set  $i\in H$, and assume, w.l.o.g., that $S_i = \sigma_t$. Any substitution of $S_i$ to $\sigma_r$ for $r=1..t-1$ eliminates all $P$-matches that contain index $i$, and Lemma \ref{EliminatingDnaClaim} implies that at most one of these substitutions can create a new $P$-match. Therefore, there are at least $t-2$ substitutions of the character $S_i$ that eliminate all $P$-matches that include $i$ and create no new $P$-matches. Thus, once a set of positions that cover all matches is identified, an eliminating set can be constructed by finding for each position $i$ in the hitting set a substitute character that does not create a new $P$-match. The argument above implies that there are at least $t-2$ substitute characters that guarantee this for every position in the hitting set. 
	\end{proof} 
	Claim \ref{corollaryHittingEqualEliminating} implies a simple algorithm for computing a minimal eliminating set in the non-binary alphabet case. The outline of such an algorithm is: 
 	\begin{algorithm}[H]
		\caption{Computing minimal eliminating set}
		\label{algBasicAlgForEliminating}
		\begin{algorithmic}[1]
			\State Compute the set of intervals $I$ corresponding to all $P$-matches in $S$.
			\State Compute a minimal hitting set $H$ for $I$.
			\State For every $j \in H$, find a substitution character $\sigma$, such that substituting $S_j$ with $\sigma$ does not create new $P$-matches.
		\end{algorithmic}
	\end{algorithm}	
	Step $1$ is implemented either by the KMP algorithm or by a suffix tree, and is achieved in $O(n+k|\Sigma|)$ (see brief review in Section \ref{TheoreticalWork}). 
	Step $2$ is implemented by a simple greedy algorithm that is described in Section \ref{HittingSetAlg} below in $O(|I|)$ time. Lastly, Step $3$ is implemented by considering an arbitrary substitute characters for every position $j\in H$ and checking the interval $[j-k+1 , j+k-1]$ for a new $P$-match. If no $P$-match is found, then this character is chosen, and if a $P$-match was found, then a different (arbitrary) substitute character is chosen (Claim \ref{corollaryHittingEqualEliminating} guarantees that at most one character can create a new $P$-match). Therefore, the time complexity of step $3$ is $O(k \cdot |I|)$. Finally, the total time complexity of Algorithm \ref{algBasicAlgForEliminating} is $O(n+ k \cdot (|I| + |\Sigma|)) = O(k \cdot n)$. 
	
	Note that this algorithm has at least $t-2$ degrees of freedom for choosing a substitute character for each position in the hitting set. However, in the binary case where $t=2$ we are not guaranteed that every hitting set can be used to generate a valid eliminating set. We address this issue in detail in Appendix \ref{EliminatingOverBin}.

	\subsection{Efficient algorithm for finding a hitting set}\label{HittingSetAlg}	

	The minimal hitting set problem we defined is a special case of the set cover problem,  which is a very well known NP-complete problem (\cite{HittingSetNpHard}), but in the special case of interval sets it has a simple linear algorithm (see \cite{GreedyAlgForHittingSet}), which we present below in Algorithm \ref{algEliminatingIntervals} for completeness. The algorithms goes through the intervals, in order, and when it encounters an interval that is not covered, it adds its right-most position to the hitting set. Assuming the intervals are already sorted, the complexity of the algorithm is $O(|I|)$ time and $O(1)$ extra space. The correctness of the algorithm is thus established by Claim \ref{claimMinHittingSetIntervals} below. 
\begin{algorithm}[t]
	\caption{Computing minimal hitting set for a set of intervals $I$}
	\label{algEliminatingIntervals}
	\begin{algorithmic}[1]
		\State Sort the intervals in $I$ in increasing order of the rightmost index they contain.
		\While{$I \ne \emptyset$}
		\State Pick the first ending interval, $[l,r] \in I$, and add position $r$ into $H$.
		\State Remove all intervals that contain position $r$ from $I$.
		\EndWhile
	\end{algorithmic}
\end{algorithm}	
	\begin{claim}\label{claimMinHittingSetIntervals}
		The set $H$ returned by Algorithm \ref{algEliminatingIntervals} is a minimal hitting set of the input set of intervals $I$.
	\end{claim}
	\begin{proof}
		The algorithm removes an interval from $I$ only if $H$ covers it, implying that $H$ is a hitting set for $I$. We are left to argue the minimality of $H$. We do this by proving that for an arbitrary hitting set $H'$ of $I$, we have $|H|<=|H'|$. Consider positions in $H$ in ascending order: $H = \{m_1,m_2,...m_l\}$. We will prove by induction on $i$ that $|H' \cap [1..m_i]| \ge i$. \\
		\textbf{Base: $i = 1$} \\
		Position $m_1$ is the rightmost position in the first ending interval in $I$. Any hitting set should cover this interval using a position that is prior to $m_1$, therefore: $|H' \cap [1..m_1]| \ge 1$.\\
		\textbf{Step:} Assume correctness of the claim for all $i'< i$ and prove for $i$.
		Let $[l,r]$ denote the interval for which the algorithm decided to add position $m_i$ to $H$ (step $3$ of the algorithm). The algorithm decided to add position $m_i$ because the interval was not covered by positions $\{m_1..m_{i-1}\}$ implying that $l > m_{i-1}$ and $r = m_i$. $H'$ is a hitting set of $I$ so it has to cover interval $[l,r]$. We get that: $$|H' \cap [1..m_i]| \ge |H'\cap[1..m_{i-1}]| + |H' \cap [l..m_i]| \ge |H'\cap[1..m_{i-1}]| + 1.$$ Since the induction hypothesis implies that $|H'\cap[1..m_{i-1}]| \ge i-1$ we get that $|H' \cap [1..m_i]| \ge i - 1 + 1 = i$, as required.
		
		Applying this inductive claim to $i=|H|$, we get that any arbitrary hitting set $H'$ of $I$ satisfies $$ |H'| >= |H'\cap [1..m_{|H|}]| >= |H|. $$
	\end{proof}

	\subsection{Proof of Lemma \ref{EliminatingDnaClaim}}\label{ProofOfEliminatingDnaClaim}
	Recall that Lemma \ref{EliminatingDnaClaim} states that if a position $j$ in $S$ belongs to a $P$-match, then substituting $S_j$ with any character can create at most one new $P$-match. The following proof follows similar lines of arguments as in the proof of lemma $18$ in \cite{deletingPatternMultiD}. 
	\begin{proof}
	Assume, in contradiction, that substituting $S_j$ creates two new P-matches. This may be either by a single substitution $S_j \leftarrow \sigma$ or by two different substitutions $S_j \leftarrow \sigma_1$ and $S_j \leftarrow \sigma_2$. Let $i$ denote the starting position of the original $P$-match and let $i_1$ and $i_2$ denote the two starting positions of the two new $P$-matches. 
	Denote by $y_1,y_2$ the offsets (in $[0,k-1]$) of the substituted position w.r.t the newly created $P$-matches, i.e, $y_1=j-i_1, y_2=j-i_2$. \\
	The fact that three $k$-long substrings starting in positions $i,i_1,$ and $i_2$ are nearly identical implies the following basic observation: for every $t \in \{1,2\}$ and every offset $x \in [0,k-1] \setminus \{y_t\}$ we have $S_{i+x}=S_{i_t+x}=P_{x+1}$ and for $y_t$ we have $S_{j}= S_{i_t+y_t} \ne S_{i+y_t}$. 
	This is because of the one exact $P$-match starting in position $i$ and the two near exact matches starting in position $i_1$ and $i_2$. 
	We use the series of equations in this basic observation to define the following undirected graph $G=(V,E)$ :
	$$V=[1,n], E=\{(u,v) | x=u-i \in [0,k-1] \wedge v\in \{i_1+x, i_2 +x\} \wedge v \ne j\}.$$
	The basic observation we stated above implies that if positions $u$ and $v$ are connected in $G$ then we have $S_u = S_v$. We will reach a contradiction by showing there is a path in $G$ from $j$ to either $i+y_1$ or $i + y_2$. Denote by $\Delta_1$ and $\Delta_2$ the distance between the starting positions of the original $P$-match and the two newly created $P$-matches: $\Delta_t =|i_t-i|$. Now, distinguish between the following two cases:\\ \\
	\underline{Case 1:} The original $P$-match is on the same side of the two newly created $P$-matches: $i<i_1$ or $i>i_2$. Assume, w.l.o.g., that $i<i_1$. (If $i>i_2$, then we can reverse the sequence $S$ and the pattern $P$ and then obtain the desired configuration with the reversed sequences.) \\
	In this case, $(u,v) \in E$ iff $u-i \in [0,k-1] \wedge v-u \in \{\Delta_1, \Delta_2\} \wedge v \ne j$. We will reach a contradiction by showing a path of length $3$ in $G$ connecting positions $j$ and $i+y_2= j- \Delta_2$. Consider the following series of positions: $j \rightarrow j + \Delta_1 \rightarrow j + \Delta_1 - \Delta_2 \rightarrow j - \Delta_2$. Notice that the first, third and fourth positions in this walk belong to the range $[i, i + k -1]$: $i \le j-\Delta_2 < j + \Delta_1 - \Delta_2 < j < i + k$. The second and the third inequalities follow from the assumption that $\Delta_2 > \Delta_1$ and that both are positive. The first and forth inequalities follow from $i_2 \le j < i + k$ (position $j$ belongs to the $k$-long substrings starting in positions $i, i_2$). This implies that the three steps in this walk correspond to edges in $G$: \\
	\begin{itemize}
		\item $(j, j+\Delta_1)\in E$ because $j-i\in [0, k-1]$ (established above), $(j+\Delta_1)-j = \Delta_1$ and $j+\Delta_1 > j$
		\item $(j+\Delta_1, j+\Delta_1 - \Delta_2)\in E$ because $(j+\Delta_1 - \Delta_2)-i \in [0,k-1]$ (established above), $(j+\Delta_1) - (j+\Delta_1 - \Delta_2) = \Delta_2$ and $j+\Delta_1 > j$
		\item $(j+\Delta_1 - \Delta_2, j-\Delta_2) \in E$ because $(j-\Delta_2) - i \in [0, k-1]$ (established above), and $(j+\Delta_1 - \Delta_2) - (j-\Delta_2) = \Delta_1$, and $j+\Delta_1-\Delta_2 < j$
	\end{itemize}
	\ \\
	\underline{Case 2:} The original $P$-match is between the two newly created matches: $i_1<i<i_2$.
	We will reach a contradiction by showing that there is a path in the graph connecting positions $j$ and $i+y_1 = j+\Delta_1$, but the length of this path will depend on the specific values of $\Delta_1$ and $\Delta_2$. 
	In this case, $(u,v) \in E$ iff $u-i \in [0,k-1] \wedge v-u \in \{-\Delta_1, \Delta_2\} \wedge v \ne j$. Consider a walk through positions that starts in position $v_0 = j$ and proceeds according to the following protocol: 
	\begin{equation*}
	v_t=\begin{cases}
	v_{t-1} - \Delta_1, & \text{If $v_{t-1} - \Delta_1 > j - \Delta_2$}\\
	v_{t-1} + \Delta_2, & \text{Otherwise}
	\end{cases}
	\end{equation*}
	Informally, the series takes backward-$\Delta_1$ steps as long as the position is greater than $j-\Delta_2$, and when it cannot, it takes a forward-$\Delta_2$ step.
	We will show that this walk reaches position $i+y_1 = j+\Delta_1$, and each step in this walk from $j$ to $j+\Delta_1$ corresponds to an undirected edge in $G$. First, note that the walk is confined to the range $[j - \Delta_2+1, j + \Delta_1]$. The lower bound directly follows from the definition of the backward step, and the upper bound follows from the fact that forward steps are taken from positions no larger than $j-\Delta_2+\Delta_1$ (otherwise a backward step is taken). Now, because the size of this range is exactly $\Delta_1+\Delta_2$, no position in the range can be approached from more than one position. Because the walk range is finite, this implies that the walk will eventually close a cycle and return to position $j$ with a backward-$\Delta_1$ step from position $j+\Delta_1$.\\
	We are left to show that all steps in this walk from $j$ to $j+\Delta_1$ correspond to edges in $G$. By design, for every $t>0$, $v_{t} - v_{t-1} \in \{-\Delta_1, \Delta_2\}$ and $v_t \ne v_0 = j$. Then, the steps in the path correspond to edges in $G$ if the range of the walk, $[j - \Delta_2+1, j + \Delta_1]$, is in $[i, i + k - 1]$. Position $j$ belongs to the near exact $P$-match starting in position $i_2$, therefore it holds that $j-\Delta_2 \ge i_2 -\Delta_2 = i_2 - (i_2 - i) = i$. Similarly, position $j$ belongs to the near exact $P$-match starting in position $i_1$, therefore it holds that $j +\Delta_1 < i_1 + k + \Delta_1 = i_1 + k + (i - i_1) = i + k$. 	
	\end{proof}		
	\section{Introducing position-specific restrictions}\label{PositionSpecificRestrictions}
	When specifying a sequence for synthesis, we will often be restricted to change the sequence only in a given set of positions. For example, if the sequence contains a coding sequence for a given gene, then we would typically wish to avoid substitutions that change the resulting sequence of amino acids. Non-coding positions may also be restricted if they fall in regulatory sequences (promoters, enhancers, etc.). There are two different ways to specify such restrictions: 
	\begin{itemize}
		\item Position-specific hard restrictions: the user provides a set of indices that are not allowed to be changed. The objective will be to clean $S$ using a minimal number of changes in the set of allowed positions. 
		\item Position-specific soft restrictions: the user specifies a penalty for a letter change in each position along the sequence. The objective here is to clean $S$ at a minimum-cost. Note that hard restrictions can be implemented in this framework by associating positions that are not allowed to be changed with a very high cost (practically $\infty$). In this section we consider cost schemes where the cost of substituting a given position does not depend on the base we substitute it with. Later, in Section \ref{FSMIntro} we consider a more general cost scheme where the cost associated with a substitution in a given position may depend on the base we substitute it with.
	\end{itemize}
	\subsection{Position-specific hard restrictions}
	Given a set of positions that are not allowed to be modified, $R$, we find a minimal elimination set by modifying step $2$ of Algorithm \ref{algBasicAlgForEliminating} to compute a minimal hitting set $H$ among hitting sets that do not intersect $R$. This is achieved by modifying step $3$ in Algorithm \ref{algEliminatingIntervals} to select the right-most position in $[l,r] \setminus R$ to add to the hitting set. Note that this modification does not influence the complexity of the algorithm, so a minimal elimination set is still computed in $O(kn)$ even under hard restrictions. We now prove that this modification yields the required outcome. 
	\begin{claim}\label{HardRestrictionIsMinimal}
		The set $H$ returned by the modified version of Algorithm \ref{algEliminatingIntervals} is a minimal hitting set of the input set of intervals $I$, among hitting sets that do not intersect the set of restricted positions $R$.
	\end{claim}
	\begin{proof}
		The proof is similar in spirit to the proof of Claim \ref{claimMinHittingSetIntervals}. $H$ is a hitting set of $I$, because the algorithm makes sure to cover all intervals. Moreover, $H$ does not intersect $R$, because the positions added to $H$ in the modified step $3$ are never in $R$. We are left to argue that every other hitting set $H'$ that does not intersect $R$ is not smaller than $H$. Consider positions in $H$ in ascending order: $H = \{m_1,m_2,...m_l\}$. We will prove by induction on $i$ that $  |H' \cap [1..m_i]| \ge i$. \\
	\textbf{Base: $i = 1$} \\
	Position $m_1$ is the rightmost position that is allowed to be changed in the first ending interval in $I$. Any valid hitting set should cover this interval using a position that is prior to $m_1$, therefore: $|H' \cap [1..m_1]| \ge 1$.\\
	\textbf{Step:} Assume correctness of the claim for all $i'< i$ and prove for $i$.
	Let $[l,r]$ denote the interval for which the algorithm decided to add position $m_i$ to $H$ (step $3$ of the modified version above). The algorithm decided to add position $m_i$ because the interval was not covered by positions $\{m_1..m_{i-1}\}$ implying that $l > m_{i-1}$. $H'$ has to cover interval $[l,r]$ using at least one position from $[l, m_i]$ because $m_i$ is the rightmost position in $[l,r]$ that is allowed to be changed. We get that: $$|H' \cap [1..m_i]| \ge |H'\cap[1..m_{i-1}]| + |H' \cap [l..m_i]| \ge |H'\cap[1..m_{i-1}]| + 1.$$ Since the induction hypothesis implies that $|H'\cap[1..m_{i-1}]| \ge i-1$ we get that $|H' \cap [1..m_i]| \ge i - 1 + 1 = i$, as required.

	Applying this inductive claim to $i=|H|$, we get that any arbitrary hitting set $H'$ of $I$ that does not intersect $R$ satisfies $$ |H'| >= |H'\cap [1..m_{|H|}]| >= |H|. $$
	\end{proof}

	\subsection{Position-specific soft restrictions}\label{PositionSpecificSoft}
	We implement position-specific soft restrictions by introducing a cost function on sequence positions. The cost function, $cost(i)$ specifies the cost incurred by substituting position $i$ such that all possible substitutions of $i$ have the same cost. Our objective is to find a minimum-cost eliminating set of a pattern $P$. 
	 As in the case of hard restrictions, we do this by modifying step $2$ of Algorithm \ref{algBasicAlgForEliminating} to compute a minimum-cost hitting set. This is done by applying a relatively straightforward dynamic programming algorithm that computes two 1D tables, $H$ and $A$. Entry $H[i]$ holds a minimum-cost hitting set for the set of all intervals in $I$ that are contained in the prefix $[1..i]$ and entry $A[i]$ holds its cost, i.e., $A[i] = cost(H[i]) = \sum\limits_{j \in H[i]} cost(j)$.
	The tables $H$ and $A$ are calculated using Algorithm \ref{algPositionDP}, described below.
	  The time complexity of the algorithm is $O(n + k \cdot I)$ because for each examined position ($i$) that ends an interval we scan the preceding $k$ indices. The extra space complexity is dominated by the dynamic programming table $H$, since its entries hold sets. In order to reduce the extra space used we can save only a pointer to the last position in $H[i]$ and use these pointers to reconstruct $H[i]$ by back tracing. Notice that this modification increases the time complexity of step $2$ in Algorithm \ref{algBasicAlgForEliminating}, but the total time complexity of Algorithm \ref{algBasicAlgForEliminating} remains the same $(O(k\cdot (n + |\Sigma|)))$.
	The correctness of the algorithm is established by the the following claim:
	 \begin{algorithm}[t]
		\caption{Computing a minimum-cost hitting set}
		\label{algPositionDP}
		\begin{algorithmic}[1]
			\State Initialization:  $H[0]= \emptyset, A[0] = 0$.
			\State Update step for index $i$:
			\NoNumber{If there is an interval ending in position $i$, then compute}
			\NoNumber{$j=\argmin\limits_{l\in [i-k+1,i]}\{A[l-k] + cost(l)\}$ and set:}
			\vspace{1mm}
			\NoNumber{~~~ $H[i] = H[j-k]\cup{j}$}
			\NoNumber{~~~ $A[i] = A[j-k] + cost(j)$}
			\NoNumber{Otherwise, set:}
			\NoNumber{~~~ $H[i] = H[i-1]$}
			\NoNumber{~~~ $A[i] = A[i-1]$}
		\end{algorithmic}
	\end{algorithm}	

	\begin{claim}
		$H[i]$ holds a minimum-cost hitting set of the set of all intervals in $I$ that are contained in the prefix $[1..i]$ and $A[i]$ holds its cost. 
	\end{claim}

	\begin{proof}
		By induction on $i$. \\
		\textbf{Base:} $i=0$: \\
		The empty prefix has an empty hitting set with cost $0$.\\ 
		\textbf{Step}: Assume correctness of the claim for all $i'< i$ and prove for $i$. 
		$H[i]$ is a hitting set for the given set of intervals because the algorithm makes sure to cover all intervals in the range $[1..i]$. We are left to argue the minimality of $H[i]$ and we establish it by proving that for an arbitrary hitting set $H'$ for the same set of intervals we have $cost(H[i]) \le cost(H')$. \\
		If there is no interval ending in position $i$, then $H[i] = H[i-1]$ and the induction hypothesis implies that $cost(H[i-1]) \le cost(H')$. Otherwise, there is an interval ending in position $i$. Let $j$ and $l$ be the rightmost indices of $H[i]$ and $H'$ that cover that interval correspondingly. The induction hypothesis implies that $cost(H' \cap [1,l-k]) \ge A[l-k]$. According to how index $j$ is set by the algorithm, $A[l-k] + cost(l) \ge A[j-k] + cost(j)$. By combining the inequalities above with the definition of $H'$ and $H[i]$ we get:
		$$cost(H') \ge cost(H' \cap [1,l-k]) + cost(l) \ge A[l-k] + cost(l) \ge A[j-k] + cost(j) = cost(H[i]).$$ 
	\end{proof}
	\section{Dynamic programming algorithms for a generalized elimination problem}\label{FSMIntro}
	In this section, we generalize the elimination problem in two directions. First, we allow the specification of multiple unwanted patterns, since usually there is more than one pattern to eliminate (e.g., multiple binding sites of different transcription factors and/or restriction enzymes). Second, we allow a more general cost scheme than the one considered in Section \ref{PositionSpecificSoft}, where the cost of substituting a given position may depend on the target base. Assuming an additive cost function, this scheme implies a cost on any sequence $S$ that has the same length ($n$) as the target sequence: $cost(S) = \sum_{i=1}^{n}cost(i,S_i)$. This generalized cost scheme allows the user to define a preference toward certain type of substitutions (e.g. transitions versus transversions), and to allow a wider range of synonymous substitutions (that do not change the encoded amino acids in a gene). Using this scheme we redefine our objective as \textbf{finding a minimum-cost sequence of length $n$ that does not contain any unwanted pattern}. Note that in this redefined objective the target sequence ($S$) is not explicitly specified, but it can be thought of as being the minimum-cost sequence of length $n$ (with possible instances of unwanted patterns).
	
	This objective cannot be solved by slight modifications to the previous algorithms because we can no longer separate the two decisions that we are making: the set of positions to substitute and the target bases we substitute to. For example, consider the following scenario, where we wish to eliminate pattern $P=ACT$ from the target sequence $S=ACACT$ using the following cost function:	
	\begin{center}
		\begin{tabular}{l l l l l l} 
			\hline
			position ($i$) & $1$ & $2$ & $3$ & $4$ & $5$ \\
			 $S[i]$ & $A$ & $C$ & $A$ & $C$ & $T$ \\ 
			\hline
 			
			$cost(i, A)$ & 0 & 2 & 0 & 3 & 3\\ 
			$cost(i, T)$ & 2 & 2 & 1 & 3 & 0\\
			$cost(i, C)$ & 2 & 0 & 4 & 0 & 3\\
			$cost(i, G$) & 2 & 1 & 4 & 3 & $\infty$\\
			\hline
			
		\end{tabular}
	\end{center}
	There is a $P$-match starting in position $3$ that should be eliminated. The minimum-cost sequence without a $P$-match is $AGTCT$ of cost $2$. Note that in this case it is beneficial to substitute two positions ($2, 3$), one of them creates a new $P$-match and the other eliminate the newly created $P$-match. The previous approach which restricts itself to substitutions that do not create new $P$-matches would substitute only one position (for example position $4$) and would result in a higher cost of $3$. Thus, a solution to this generalized elimination problem requires an algorithm that jointly considers the substituted positions and the bases we choose to substitute to.  
	
	 To solve this problem, we suggest a simple dynamic programming algorithm based on a finite state machine (FSM) that generates all (and only) sequences without unwanted patterns. Given such an FSM, Algorithm \ref{algNaiveFsm} below finds the minimum-cost sequence of a given length that the FSM generates. This implies that the elimination problem reduces to finding such an FSM, which is what we do in Sections \ref{NaiveFsm} and \ref{KmpFsm}.
	\begin{definition}
		An FSM that generates sequences is defined by the tuple $(\Sigma, V, f)$ where
		\begin{itemize}
			\item $\Sigma$ is the alphabet of the generated sequences.
			\item $V$ is the state space which includes a single initial state $v_{init} \in V$.
			\item $f:V \times \Sigma \rightarrow V$ is a partial transition function (i.e, not defined for all $(v, \sigma) \in V \times \Sigma$).
		\end{itemize}
	\end{definition}
	 A sequence $S$ of length $n$ is said to be generated by a given $\FSM$ if there is a path through states of the $FSM$ $v_{init} = v_0, v_1 ... , v_n$ such that $f(v_{i-1}, s_i) = v_{i} \ \ \forall i\in [1..n]$. 
	Note that because the transition function $f$ is partial, then not all sequences have a generating path. Furthermore, because the FSM is deterministic and has a single initial state, then the generating path is unique, and we denote by $FSM(S)$ the final state ($v_n$) in that path.
	
	We can find the minimum-cost sequence of a given length generated by the $\FSM$ by a rather straightforward calculation of a dynamic programming table $A$ s.t $A[i,v]$ holds the minimum cost of a sequence $S$ of length $i$ that is generated by the $\FSM$ and $\FSM(S) = v\in V$. 
	Note that this algorithm does not involve an initial step of finding all pattern matches in the target sequence. This is because it considers all clean sequences in parallel and does not start from a specific target sequence, as the algorithms in sections \ref{ConnectionBetweenElimAndHitting} and \ref{PositionSpecificRestrictions} did.
 	\begin{algorithm}[H]
	\caption{A dynamic programming algorithm for finding the minimum-cost sequence of length $n$ generated by a given FSM = $(\Sigma,V,f)$}
	\label{algNaiveFsm}
		\begin{algorithmic}
		\State \textbf{Initialization:}
		\NoNumber{ $A[0,v]=\begin{cases}
			0, & \text{if }v = v_{init}\\
			\infty, & \text{otherwise}
			\end{cases}$ 
		}
		\State \textbf{Update:} 
		\NoNumber{For all $i=1..n, v\in V$:}
		\NoNumber{$A[i,v]=\min\limits_{u,\sigma: f(u, \sigma) = v} \{A[i-1, u] + cost(i, \sigma)\}$\\}
		\NoNumber{$A^*[i,v] = \argmin\limits_{u,\sigma: f(u, \sigma) = v} \{A[i-1, u] + cost(i, \sigma)\}$}
		\State \textbf{Constructing $S$}: 
		\NoNumber{$i=n, \ v_n=\argmin\limits_{u \in V} \ A[n,u]$}
		\NoNumber{For all $i=n..1$: $(v_{i-1}, S_i) = A^*[i,v_i]$}
	\end{algorithmic}
	\end{algorithm}	
	\begin{claim}
		$A[i,v]$ holds the minimum cost of a sequence $S$ of length $i$ that is generated by the $\FSM$ s.t $\FSM(S)=v$
	\end{claim}
	\begin{proof}
		By induction on $i$: \\
		\textbf{Base:} $i=0$: \\
		The only sequence of length $0$ is $\varepsilon$ and it holds that $\FSM(\varepsilon) = v$ iff $v = v_{init}$.\\
		\textbf{Step}: \\
	Assume correctness of the claim for all $i'< i$ and all $v \in V$, and prove for $i$ and an arbitrary $v\in V$. \\
	We first prove that $A[i,v] \leq cost(S)$ for any sequence $S$ of length $i$ that is generated by the $\FSM$ s.t $\FSM(S) = v$. Let $S$ be such a sequence and let $\sigma = S_i$, then $S=S'\sigma$, and let $u$ be the state such that $\FSM(S') = u$. Thus, $f(u, \sigma) = v$ and the induction hypothesis implies that $A[i-1, u] \leq cost(S')$. Thus, using the update step definition we get that $$A[i,v] \leq A[i-1, u] + cost(i, \sigma) \leq cost(S') + cost(i,\sigma) = cost(S).$$
	
	We are left to show that there is a sequence $S$ of length $i$ that is generated by the $\FSM$ s.t $FSM(S) = v$ and $cost(S) = A[i,v]$. Let $(u, \sigma)$ be the pair that minimizes the update step, meaning that $f(u, \sigma) = v$ and $A[i,v]=A[i-1,u]+cost(i,\sigma)$. The induction hypothesis implies that there is a sequence $S'$ of length $i-1$ that is generated by the $\FSM$ s.t $\FSM(S') = u$ and $A[i-1, u] = cost(S')$. Then, $S=S'\sigma$ is of length $i$, is generated by the $\FSM$, and $\FSM(S) = f(u,\sigma) = v$. This gives us $$cost(S)=cost(S') + cost(i, \sigma) = A[i-1,u] + cost(i, \sigma) = A[i,v].$$
	\end{proof}
 	\textbf{Complexity:} The space complexity of storing the dynamic programming tables $A$ and $A^*$ is $O(n \ |V|)$. Adding the space complexity required for holding the transition function for the FSM $(|f|)$, we get that the total extra space complexity is $O(|f| + n \ |V|)$. Note that $|V|\cdot |\Sigma|$ is an upper bound for $|f|$. The time complexity of the update of cell $A[i,v]$ is linear in the size of the source set for state $v$: $\{(u, \sigma) \ |  \ f(u,\sigma) = v\}$. Assuming that the source sets of all states are specified in the input given to the algorithm, the total time complexity for updating all cells in the $i^{th}$ row of the table $(A[i,])$ is the sum of the sizes of all source sets. The source sets of the states in $V$ forms a disjoint partition of the Cartesian product $V \times \Sigma$, and therefore the total time complexity for updating every row of the matrix is at most $|V|\cdot |\Sigma|$ (which is also an upper bound of the size of the FSM). In conclusion, the total time complexity is $O(n \ |V| \ |\Sigma|)$. 
 	
 	  In the following two subsections, we show a couple of FSMs that generate all (and only) sequences without unwanted patterns and show how to compute the source sets for each one of them. 
	\subsection{A naive FSM based on the de Brujin graph}\label{NaiveFsm}
	The first FSM we suggest for this purpose is based on the de Brujin graph \cite{DeBruijn}. Let $\mathcal{P}$ be a collection of unwanted patterns and let $k$ be an upper bound on their length. The de Bruijn-inspired FSM for generating clean sequences is denoted by $DB_{\mathcal{P}}$ and defined as follows: $V$ corresponds to the set of all $k$-long $\mathcal{P}$-clean sequences, and the transition function $f(v,\sigma)$ is defined by computing the $k$-long suffix of $v\sigma$ (adding $\sigma$ to $v$ and removing its first character). Importantly, $f(v,\sigma)$ is defined only if this $k$-long suffix corresponds to a state in $V$. Furthermore, in this FSM, we deviate from the requirement of having a single initial state by allowing every state to be an initial state, and letting the first state define the first $k$ characters of the generated sequence. Note that despite having more than one initial state, a sequence $S$ that is generated by $DB_{\mathcal{P}}$ has only one path through the states: $v_k, ... v_n$ such that $v_i = S_{i-k+1..i}$ and $f(v_{i-1}, s_i) = v_{i} \ \ \text{for every } i\in [(k+1)..n]$. Therefore, $DB_{\mathcal{P}}(S)$, the final state generating a given sequence, $S$, in this FSM, $DB_{\mathcal{P}}$, is well defined.
\begin{claim}
$DB_{\mathcal{P}}$ generates all and only sequences (of length at least $k$) without unwanted patterns from $\mathcal{P}$.
\end{claim}
\begin{proof}
By induction on $i$, the length of the sequence: \\
\textbf{Base:} $i=k$: \\
Following the definition of $DB_{\mathcal{P}}$, all (and only) $k$-long $\mathcal{P}$-clean sequences are initial states of $DB_{\mathcal{P}}$. \\
\textbf{Step}: \\
Assume correctness of the claim for all $i' < i$ and prove for $i$. Let $S$ be a sequence of length $i$ that is $\mathcal{P}$-clean, then $S=S'\sigma$ such that $S'$ is a $\mathcal{P}$-clean sequence of length $i-1$. Using the induction hypothesis, $S'$ is generated by $DB_{\mathcal{P}}$. Let $DB_{\mathcal{P}}(S') = u$. The $k$-long suffix of $u\sigma$ is also $\mathcal{P}$-clean, therefore $f(u, \sigma)$ is defined, meaning that $S$ is generated by $DB_{\mathcal{P}}$ and it holds that $DB_{\mathcal{P}}(S) = f(u, \sigma)$. \\
 We are left to show that $DB_{\mathcal{P}}$ generates only sequences without unwanted patterns. Let $S$ be a sequence of length $i$ generated by $DB_{\mathcal{P}}$ and let $\sigma = S_i$ then $S = S' \sigma$. Using the induction hypothesis, $S'$ is of length $i-1$ and is generated by $DB_{\mathcal{P}}$ and therefore does not contain an unwanted pattern. Adding $\sigma$ at the end of $S'$ does not introduce a $\mathcal{P}$-match because the $k$-long suffix of $S$ corresponds to a state in $V$. 
\end{proof}
The size of the state space of this FSM is very large ($\Omega(\Sigma^k \setminus \mathcal{P})$), and it dominates the complexity of using this $FSM$ in the context of Algorithm \ref{algNaiveFsm}. We therefore turn to look for a significantly smaller FSM that serves the same purpose.
\subsection{A smaller KMP-based FSM}\label{KmpFsm}
To produce a smaller FSM for this problem, we utilize the KMP-inspired automaton suggested by Aho and Corasick \cite{KMPBasedPatternMatching} (see brief review in Section \ref{TheoreticalWork}). Recall that this automaton finds all matches of a set of patterns by keeping track of the longest suffix of the traced sequence that is also a prefix of a given pattern. We extend this FSM to avoid complete matches. This approach will let us generate all and only sequences without unwanted patterns. 

 We denote the KMP-inspired FSM for a given collection $\mathcal{P}$ of unwanted patterns by $KMP_{\mathcal{P}}$ and define it as follows: we first define $\prefP$ as the set: $\{w \ | \ \exists P\in \mathcal{P} \text{ s.t. } w \text{ is a prefix of }P\}$. Then $V = \prefP \setminus \{w \ | \exists P \in \mathcal{P} \text{ s.t. } P \text{ is a suffix of } w\}$. In other words, there is a state for every prefix of a pattern in $\mathcal{P}$ that does not end with an unwanted pattern. We designate the state corresponding to the empty string, $\varepsilon$, as the initial state $v_{init}$. The transition function $f(v, \sigma)$ is defined as follows: if there is a suffix of $v\sigma$ that is an unwanted pattern, then $f(v, \sigma)$ is not defined. Otherwise, $f(v, \sigma)$ is the \emph{longest suffix} of $v\sigma$ that is in $\prefP$.
\begin{claim}
	$KMP_{\mathcal{P}}$ generates all and only sequences without unwanted patterns from $\mathcal{P}$.
\end{claim}
\begin{proof}
	We first prove that any $\mathcal{P}$-clean sequence $S$ can be generated by the FSM by induction on the length of $S$. For length $0$, the only $\mathcal{P}$-clean sequence is $\varepsilon$, which is generated by $KMP_{\mathcal{P}}$ and $KMP_{\mathcal{P}}(\varepsilon) = v_{init}$. For longer $S$, there is a sequence $S'$ such that $S=S'\sigma$. The induction hypothesis implies that $S'$ is generated by $KMP_{\mathcal{P}}$. Let $KMP_{\mathcal{P}}(S') = u$, then the sequence $u\sigma$ is $\mathcal{P}$-clean because it is a suffix of $S$, implying that $f(u, \sigma)$ is defined and is equal to $v$. Thus, $S$ is generated using the path that generates $S'$ appended by state $v=f(u,\sigma)$. 
	
	For the opposite direction we need to strengthen the induction hypothesis and show that every generated sequence, $S$, is $\mathcal{P}$-clean and that the state $KMP_{\mathcal{P}}(S)$ corresponds to the longest prefix in $\prefP$ that is also a suffix of $S$. For length $0$, the only generated sequence is $\varepsilon$, which is $\mathcal{P}$-clean and $KMP_{\mathcal{P}}(\varepsilon) = v_{init}$, which is the longest prefix in $\prefP$ that is also a suffix of $\varepsilon$. For longer $S$, there is a sequence $S'$ such that $S=S'\sigma$. The induction hypothesis implies that $S'$ is $\mathcal{P}$-clean and $KMP_{\mathcal{P}}(S') = u$ corresponds to the longest prefix in $\prefP$ that is also a suffix of $S'$.	The definition of the transition function $f$ implies that $v=f(u,\sigma)$ is the longest prefix in $\prefP$ that is a suffix of $u\sigma$. Because $u\sigma$ is a suffix of $S$, then so is $v$. We are left to prove that any longer suffix of $S$, $w$, is not a prefix in $\prefP$. If $w$ is shorter than $u\sigma$, then it is not in $\prefP$, because of the way the transition function is defined. If, on the other hand, $w$ is longer than $u\sigma$, then $w=x\sigma$, and $x$ is a suffix of $S'$. The induction hypothesis implies that $u$ is the longest suffix of $S'$ that is in $\prefP$, and $x$ is longer than $u$, so it cannot be in $\prefP$. In conclusion, $v$ is the longest prefix in $\prefP$ that is a suffix of $S$, and since $v\in V$, then $S$ does not have a suffix that is a $\mathcal{P}$-match. Since its prefix $S'$ is $\mathcal{P}$-clean, then $S$ itself is also $\mathcal{P}$-clean. 
	\end{proof}
The size of the state space of this FSM is $O(|\prefP|)$ which is significantly smaller than the size of the state space of the naive FSM described in Section \ref{NaiveFsm} ($\Omega(\Sigma^k \setminus \mathcal{P})$). Thus, by using $KMP_{\mathcal{P}}$, Algorithm \ref{algNaiveFsm} finds a minimum-cost $\mathcal{P}$-clean sequence of length $n$ in time $O(n \cdot |\Sigma| \cdot |\prefP|)$, which is linear in the size of the input. 
However, this requires an additional preprocessing step for computing $KMP_{\mathcal{P}}$. We describe the calculation of $KMP_{\mathcal{P}}$ in the Section \ref{KMPCalc} below and show that the preprocessing time and space complexity is $O(|\prefP| \cdot |\Sigma|)$. 
\subsubsection{An efficient algorithm for computing $KMP_{\mathcal{P}}$}\label{KMPCalc}
\newcommand{\PP}{\mathcal{P}}

In this section, we describe an efficient procedure for calculating the $KMP_\PP$ FSM (Aho and Corasick \cite{KMPBasedPatternMatching} describe an efficient procedure for calculating a similar FSM to $KMP_\PP$ which does not forbid pattern matches). Throughout the discussion below, we assume that the empty word $\epsilon$ is not an unwanted pattern in $\PP$. If $\epsilon\in\PP$, then $KMP_\PP$ is empty by definition and there is no sequence that does not contain unwanted patterns. To compute this FSM, we need to: 
\begin{itemize}
	\item Compute its state space  $V = \{w\in\prefP | w  \text{  does not have a suffix in }\PP\}$.
	\item Compute the (partial) transition function $f$ for every $(v,\sigma)\in V\times\Sigma$. Recall that if $v\sigma$ has a suffix in $\PP$, then $f(v, \sigma)$ is not defined. Otherwise, $f(v, \sigma)$ is the \emph{longest suffix} of $v\sigma$ that is in $V$.
\end{itemize}
Computing $V$ and $f$ requires scanning words in $\prefP\times\Sigma$ for suffixes in $\prefP$. Thus, a naive implementation would take at least quadratic time. In order to achieve this in linear time, we employ a technique originally suggested in \cite{KMP} for the construction of the KMP automaton for matching a single pattern. Our algorithm extends this technique to multiple patterns and uses it also to identify invalid transitions (which was not needed in the original pattern matching problem). The technique suggested in \cite{KMP} makes use of the auxiliary function ($g$) defined below:

\begin{definition}
	Given a collection of unwanted patterns $\PP$ and a word $w\in\Sigma^*$, we define $g(w)$ as the longest \emph{proper suffix} of $w$ that is in $\prefP$. A \emph{proper suffix} in this context is any suffix that is not equal to the entire word $w$.
\end{definition}

The relationship between the auxiliary function $g$ and the transition function $f$ is established by the following claim:

\begin{claim}\label{claim:kmp_f_g}
	Consider $(v,\sigma)\in V\times\Sigma$ s.t. $v\sigma$ does not have a suffix in $\PP$. The following two relationships hold:
	\begin{enumerate}
		\item If $v\sigma \notin \prefP$, then $f(v,\sigma)=g(v\sigma)$.
		\item $g(v\sigma) = f(g(v),\sigma)$.
	\end{enumerate}
\end{claim}
\begin{proof}
	First, note that under the conditions of the claim, the transitions $f(v,\sigma)$ and $f(g(v),\sigma)$ are defined ($v \sigma$ and $g(v) \sigma$ do not have a suffix in $\PP$). If $v\sigma \notin \prefP$, then $f(v,\sigma)\neq v\sigma$, implying that $f(v,\sigma)$ is a proper suffix of $v\sigma$. Hence, both $f(v,\sigma)$ and $g(v\sigma)$ are equal to the longest proper suffix of $v\sigma$ that is in $\prefP$, establishing (1) above.
	
	To prove (2), we need to show that $f(g(v),\sigma)$ is the \emph{longest} proper suffix of $v\sigma$ that is in $\prefP$. The definition of $f$ implies that  $f(g(v),\sigma)\in\prefP$. Furthermore, $f(g(v),\sigma)$ is a proper suffix of $v\sigma$ because $f(g(v),\sigma)$ is a suffix of $g(v)\sigma$ and $g(v)$ is a proper suffix of $v$. We are left to show that for any proper suffix $w$ of $v\sigma$ that is in $\prefP$ it holds that $|w|\leq|f(g(v),\sigma)|$. If $w=\epsilon$, then $|w|=0\leq|f(g(v),\sigma)|$. Otherwise, $w=u\sigma$, where $u$ is a proper suffix of $v$. Since $w$ is in $\prefP$, then so is $u$. So, the definition of $g$ implies that $u$ is also a suffix of $g(v)$, which implies in turn that $w=u\sigma$ is a suffix of $g(v)\sigma$. Finally, since $f(g(v),\sigma)$ is the longest suffix of $g(v)\sigma$ that is in $\prefP$, we get $|w|\leq|f(g(v),\sigma)|$, as required.
\end{proof}

The two equations in Claim \ref{claim:kmp_f_g} imply a recursive procedure for jointly computing the functions $f$ and $g$. The validity of the recursion is guaranteed by the fact that $g(v)$ is strictly shorter than $v$. The recursion halts either when $v\sigma\in\prefP$ (and then $f(v,\sigma)=v\sigma$), or when $v=\epsilon$ (and then $g(v\sigma)=\epsilon$, since the only proper suffix of $v\sigma=\sigma$ is $\epsilon$). A similar recursive procedure can also be used to compute the state space $V$ by applying the following claim:

\begin{claim}\label{claim:invalid_states}
	$v\sigma$ has a suffix in $\PP$ iff $v\sigma\in \PP$ or $g(v)\sigma$ has a suffix in $\PP$. 
\end{claim}

\begin{proof}
	If $v\sigma\in \PP$, then clearly $v\sigma$ has a suffix in $\PP$. Furthermore, since $g(v)$ is a suffix of $v$, then $g(v)\sigma$ is a suffix of $v\sigma$, and so if $g(v)\sigma$ has a suffix in $\PP$, then so does $v\sigma$. This establishes the $\Leftarrow$ direction of the claim. To establish the other direction, we consider $v\sigma\notin\PP$ s.t. $v\sigma$ has a suffix $w\in\PP$, and we show that $w$ is also a suffix of $g(v)\sigma$. We know that $w\neq\epsilon$ (because $\epsilon\notin\PP$), and that $w\neq v\sigma$ (because $v\sigma\notin \PP$). So, $w$ is a proper suffix of $v\sigma$ of the form $w=u\sigma$, where $u$ is a proper suffix of $v$. Since $u\in \prefP$ and $g(v)$ is the longest proper suffix of $v$ in $\prefP$, then $|g(v)|\geq |u|$. This implies that $u$ is a suffix of $g(v)$, because they are both suffixes of $v$, and so $w=u\sigma$ is a suffix of $g(v)\sigma$ that belongs to $\PP$.
\end{proof}
Algorithm \ref{algCalcKmpFsm} described below implements the two recursive procedures for computing $V$ and $f$ using forward recursion (establishing the base cases first). The algorithm keeps track of undefined transitions $f(v,\sigma)$ (when $v\sigma$ has a suffix in $\PP$) by setting their values to \texttt{NULL}.
The first phase of the algorithm (lines 1--6) computes all the transitions $f(v,\sigma)$ associated with elongations of pattern prefixes (where $v\sigma\in \prefP\setminus\PP$), and identifies elongations that result in complete patterns as invalid transitions (where $v\sigma\in\PP$). Note that some prefix elongations may later be identified as invalid transitions, when $v\sigma$ has a proper suffix in $\PP$.

\begin{algorithm}[b!]
	\caption{Calculating the state space $V$ and the functions $f$ and $g$}
	\label{algCalcKmpFsm}
	\begin{algorithmic}[1]
		\For {$p \in \mathcal{P}$}
		\For {$j \in [1..|p|-1]$}
		\State Set $f(p_{1..j-1}, p_{j}) \leftarrow p_{1..j}$
		\Comment\emph{prefix elongation}
		\EndFor
		
		\State Set $f(p_{1..|p|-1},p_{|p|}) \leftarrow \texttt{NULL}$
		\Comment\emph{invalid transition into complete pattern}
		\EndFor
		
		\vspace{2mm}
		
		\State $InitEmptyQueue(stateQueue)$
		\Comment\emph{initialize processing queue}
		\vspace{2mm}
		
		\State $V\leftarrow \{\epsilon\}$
		\Comment\emph{ process initial state of FSM ($\epsilon$)}
		\For {$\sigma \in \Sigma$}
		\If{ $f(\epsilon,\sigma)$ is not set yet}
		\Comment\emph{``failure" transition}
		\State Set $f(\epsilon, \sigma) \leftarrow \epsilon$
		\EndIf
		\If{$f(\epsilon,\sigma) == \sigma$}
		\Comment\emph{$\sigma$ is an FSM state}
		\State Set $g(\sigma) \leftarrow \epsilon$
		\State $stateQueue.push(\sigma)$
		\EndIf
		\EndFor
		
		\vspace{2mm}
		\While {$stateQueue$ is not empty}
		\State $v\leftarrow stateQueue.pop()$
		\State $V\leftarrow V\cup \{v\}$
		\For {$\sigma \in \Sigma$}
		\If{ $f(g(v),\sigma) == \texttt{NULL}$ }
		\Comment\emph{invalid transition}
		\State Set $f(v, \sigma) \leftarrow \texttt{NULL}$
		\EndIf
		\If{ $f(v,\sigma)$ is not set yet}
		\Comment\emph{``failure" transition}
		\State Set $f(v, \sigma) \leftarrow f(g(v),\sigma)$
		\EndIf
		\If{$f(v,\sigma) = v\sigma$}
		\Comment\emph{$v\sigma$ is an FSM state}
		\State Set $g(v\sigma) \leftarrow f(g(v), \sigma)$
		\State $stateQueue.push(v\sigma)$
		\EndIf
		\EndFor
		\EndWhile
	\end{algorithmic}
\end{algorithm}

After the initial phase, the state space $V$ is initialized with the the initial state $\epsilon$, and all transitions $f(\epsilon,\sigma)$ are considered (lines 9-17). If $f(\epsilon,\sigma)$ has not been set in the first phase of the algorithm, then no pattern in $\PP$ starts with $\sigma$, implying that $f(\epsilon,\sigma)=\epsilon$. If, on the other hand, $f(\epsilon,\sigma)$ was set in the first phase of the algorithm to $\sigma$, then there is a pattern in $\PP$ that starts with $\sigma$ and there is no pattern equal to $\sigma$, so $\sigma$ is added to the processing queue of states, and we compute $g(\sigma)=\epsilon$. 
When the second phase is complete (line 17), the processing queue contains all states in $V$ of length 1, and each of these state is associated with the correct value of $g$.

The final phase of the algorithm (lines 18-33) processes all states in $V$ using a queue that effectively implements a breadth-first search on the graph of the FSM from the initial state $\epsilon$. When $v$ is processed, the state $g(v)$ is known (because $g(v)$ is set before pushing $v$ into the queue). Furthermore, because $g(v)$ corresponds to a shorter string than $v$, it precedes it in the search order, and we are guaranteed that all transitions $f(g(v),\sigma)$ are set when $v$ is processed. If $f(g(v),\sigma)$ is undefined (set to \texttt{NULL}), we know that $g(v)\sigma$ and $v\sigma$ have a suffix in $\PP$, so $f(v,\sigma)$ should also be undefined. Note that $v\sigma$ could be a prefix of a pattern in $\PP$, and then $f(v,\sigma)$ is first defined as a prefix elongation in line 3 and later identified as an invalid transition and set to \texttt{NULL} in line 23.
Also note that if $f(g(v),\sigma)$ is defined, the algorithm ensures that $f(v,\sigma)$ will also be defined as long as it has not been set to \texttt{NULL} in the first phase (line 5).
This follows from Claim \ref{claim:invalid_states}, which implies that if $g(v)\sigma$ does not have a suffix in $\PP$ and $v\sigma$ is not a complete pattern, then $v\sigma$ does not have a suffix in $\PP$.
If $f(v,\sigma)$ has not been set in the first phase, then $v\sigma$ is not a prefix of a pattern in $\PP$, and Claim \ref{claim:kmp_f_g} is invoked to set $f(v,\sigma)$.
If, on the other hand,  $f(v,\sigma)$ was set in the first phase, then $v\sigma$ is a prefix of a pattern in $\PP$ that does not have a suffix in $\PP$.
Thus, the elongation transition is maintained, $v\sigma$ is added to the processing queue, and $g(v\sigma)$ is computed according to Claim \ref{claim:kmp_f_g}.

This procedure guarantees to process all prefixes in $\prefP$ that do not contain complete pattern matches. States in $V$ not covered by this procedure correspond to prefixes that contain unwanted patterns as non-suffix subsequences. These states are unreachable from the initial state, $\epsilon$, and are thus effectively not part of the $KMP_\PP$ FSM.
The algorithm processes every prefix in $\prefP$ once in the first phase, and the main processing loop processes each combination $(v,\sigma)\in \prefP\times\Sigma$ once. Furthermore, every calculation step done by the algorithm can be achieved in $O(1)$ as long as previously computed values of $f$ and $g$ can be retrieved in $O(1)$. Thus, the total time and space complexity of Algorithm \ref{algCalcKmpFsm} is $O(|\prefP||\Sigma|)$, meaning that it is linear in the size of the resulting FSM.

	\section{Input specification for design}
	We implemented the dynamic programming algorithm that uses the KMP-based FSM to generate all and only sequences without unwanted patterns, and an application that utilizes the algorithm for easy elimination of DNA patterns. The implementation is available in a public repository:\\ \href{https://github.com/zehavitc/EliminatingDNAPatterns.git}{https://github.com/zehavitc/EliminatingDNAPatterns.git}. For usability, we changed the inputs presented in Section \ref{FSMIntro} such that the application inputs are: 
	\begin{itemize}
		\item Sequence file - contains a raw DNA sequence: lower-case letters indicate positions that are allowed to be changed, and upper-case letters indicate positions that are not allowed to be changed. We use $IUPAC$ standard letters to indicate ambiguity in base specification (see Table \ref{IUPACTable}).
		\item Patterns file - contains a comma-separated list of patterns to eliminate. We use $IUPAC$ standard letters to indicate ambiguity in base specification (see Table \ref{IUPACTable}).
		\item Optional result file - the path to which the result should be written, if not specified, the result will be printed to the console. 
		\item Optional cost unit - the cost of substituting a letter. The default is $1$. 
		\item Optional transition transversion ratio - the cost of a letter substitution that results in a transversion (i.e., $\{A,G\}\leftrightarrow \{C,T\}$) is defined as  $cost \textunderscore unit \times  transition \textunderscore transversion \textunderscore ratio$. The default ratio is $1$. 
	\end{itemize} 
	\subsection{IUPAC support}
	The following table describe the IUPAC code:
	\begin{table}[ht]
		\caption{IUPAC code}\label{IUPACTable}
	\begin{center}
		\begin{tabular}{l l }
			\hline
			IUPAC letter & Matching bases   \\
			\hline
			$A$ & $A$   \\ 
			$C$ & $C$  \\
			$G$ & $G$  \\
			$T$ & $T$  \\
			$U$ & $T$  \\
			$R$ & $A, G$ \\
			$Y$ & $C, T$ \\
			$S$ & $G, C$ \\
			$W$ & $A, T$ \\
			$K$ & $G, T$ \\
			$M$ & $A, C$ \\
			$B$ & $C, G, T$ \\
			$D$ & $A, G, T$ \\
			$H$ & $A, C, T$ \\
			$V$ & $A, C, G$ \\
			\hline
			
		\end{tabular}
	\end{center}
\end{table}
	$IUPAC$ code in the input sequence is supported such that $cost(i, \sigma) = 0$ iff $\sigma \in IUPAC(S[i])$. For example, consider the sequence $S=rmtGD$. Let the cost unit be $1$ and the transition transversion ratio be $2$. Then we get the following cost function:
		\begin{center}
		\begin{tabular}{l l l l l l} 
			\hline
			position ($i$) & $1$ & $2$ & $3$ & $4$ & $5$  \\
			\hline
			$S[i]$ & $r$ & $m$ & $t$ & $G$ & $D$  \\ 
			\hline
			$cost(i, A)$ & $0$ & $0$ & $2$ & $\infty$ & $0$   \\ 
			$cost(i, T)$ & $2$ & $1$ & $0$ & $\infty$ & $0$  \\
			$cost(i, C)$ & $2$ & $0$ & $1$ & $\infty$ & $\infty$  \\
			$cost(i, G$) & $0$ & $1$ & $2$ & $0$ & $0$  \\
			\hline
		\end{tabular}
	\end{center}
	 The IUPAC code of $r$, in position $1$, is associated with $\{a,g\}$, implying that bases $A,G$ are associated with a zero-cost substitution, and bases $C,T$ are associated with cost of $2$, because they require a transversion-type substitution (from either $A$ or $G$). On the other hand, the IUPAC code $m$ in position $2$ means that its bases $A,C$ are associated with a zero-cost substitution, but the other two bases $(T,G)$ are associated with cost of $1$, because they can be obtained by transition-type substitutions $(C\rightarrow T or A \rightarrow G)$. Positions $4$ and $5$ are not allowed to be changed and therefore the cost of any substitution that is not in the $IUPAC$ matching bases is $\infty$. 
	
	When used in one of the unwanted patterns, $IUPAC$ code is supported such that the application replaces the given pattern with all of the patterns implied by the $IUPAC$ code. For example, consider the pattern $P=RMT$, then the application will replace it with the following set of patterns: $AAT, ACT, GAT, GCT$.
	\section{Summary and conclusion}\label{Summary}
	In this work, we suggested a systematic approach for eliminating unwanted patterns. We first established the connection between the elimination problem and the hitting set problem. We used this connection to present three linear-time algorithms that solve the problem of eliminating a single unwanted pattern, $P$, from a sequence $S$. The first two algorithms use a greedy algorithm to find a minimal hitting set with a slight computational addition that finds the substituting letter for each position in the set. This addition does not add much to the total complexity of finding a hitting set. The third algorithm supports position-specific restrictions modeled using a cost scheme that defines a cost for each substituted position. Therefore, a minimum-cost hitting set should have been found. We suggested solving this using a dynamic programming approach with linear time complexity ($O(|P||S|)$). We then generalized this approach in two directions: first to support eliminating multiple unwanted patterns, and second to support a more generalized cost scheme, where the cost of a letter substitution depends on the letter we substitute to. We described Algorithm \ref{algNaiveFsm} that solves this more general problem using a FSM that generates all and only sequences without unwanted patterns. Using this approach, the algorithm does not seek pattern matches, but generates the desired sequence from scratch. Finally, we showed an efficient FSM that can be used in Algorithm \ref{algNaiveFsm} such that the total time complexity is linear in the product of the desired sequence length and the sum of the lengths of all unwanted patterns. 
	
	Our approach to the elimination problem is strict. Our algorithm either eliminates \emph{all} instances of unwanted patterns, or reports that there is no solution (the minimum-cost clean sequence has an infinite cost). The other objectives are treated as secondary optimization tasks. As opposed to this approach, other related theoretical works treat the elimination problem as a minimization problem. For example, the problem of minimizing the number of unwanted patterns in a sequence presented in \cite{BetterPhages} has been proved to be NP-complete. Our algorithm can detect if the minimal number of unwanted patterns is zero or not, and if it is, we can find the resulting sequence efficiently. Moreover, our algorithm can be used as a subroutine in the minimization problem using a hierarchical grouping of the unwanted patterns. In this approach, each unwanted pattern is assigned with a rank that describes the priority for its removal. If there is no valid set of substitutions that eliminates all unwanted patterns, patterns can be iteratively removed from the set according to their rank, to relax the elimination constraints, until a valid (optimal) elimination set is found.
	
	The approach we suggest here has the potential to solve some of the problems with existing DNA design tools (see Section \ref{DesignTools}). One of the problems observed in existing design tools is that they do not have a well-defined behavior when posed with conflicting design requirements. When posed with such conflicting design objectives, the dynamic programming algorithm (Algorithm \ref{algNaiveFsm}), will indicate that the minimum-cost sequence has an infinite cost, and there is no finite cost solution. Furthermore, the suggested cost scheme can be used to define the constraints flexibly. One possible usage is to prioritize substitutions, such that the cost captures the expected change in the functional consequence. For example, one can set a low cost for substitutions that do not change the amino acid translation and a higher cost to substitutions that change the amino acid to a different amino acid with similar chemical properties.	The cost scheme can also be used to optimize codon usage. The codon set for the great majority of amino acids can be specified by fixed bases in the first two positions and a choice for the third position base. The cost of substituting the third base can be associated with $-log(p)$ where $p$ is the frequency of this codon. This way, the score of a sequence is inversely correlated with its likelihood under a simple codon frequency model, and a minimum-cost corresponds to maximum-likelihood. For example, Phenylalanine codons are $TTT, TTC$ so the cost of substituting the first two bases ($T$) will be set to infinity, and the cost of substituting the third base will be set infinity if substituting to $G$ or $A$, $-log(p(TTT))$ if substituting to $T$ and $-log(p(TTC))$ if substituting to $C$.  
	Another common objective of design tools is to set a $GC$ content objective. We can use the cost scheme to favor substitutions of $C$ with $G$ and $A$ with $T$ to minimize this also. 
	
	There are several key extensions that we suggest as future work. The approach described above for modeling codon usage does not work for Leucine (LEU), Arginine (ARG), and Serine (SER). Each of these amino acid has six codons, such that the first two positions cannot be fixed, and the allowed substitutions for the third base depend on the first two bases. Therefore, to fully support codon usage modeling, a fairly modest extension of the cost function needs to be defined in the context of base triplets. With this extension, one can also easily allow substituting amino acid with a different but similar amino acid. Another observation is that the unwanted patterns associated with many binding sites (e.g., transcription factor binding sites) can be represented using a short sequence with wildcard characters. Note that the number of unwanted patterns implied by a sequence with wildcard is exponential in the number of wildcard characters. 
	An interesting open question is whether there is an algorithm, which is linear in the total length of unwanted {\em wildcard} patterns, and not just in the total length of all implied patterns. Since Algorithm 
	\ref{algNaiveFsm} makes use of a FSM, it seems reasonable that one can create a FSM that recognizes this short sequence with wildcard characters and use it to eliminate the patterns. In conclusion, we made the first step in presenting a formal description of the pattern elimination problem. The  algorithms we suggest here are very efficient and relatively simple, and thus can easily be incorporated in DNA design tools. The next step in this line of research would be to extend the basic framework we propose here to allow addressing a combination of complex design objectives. 

   \section*{Acknowledgements}
    The authors wish to thank Omri Ben-Eliezer, Simon Korman, and Daniel Reichman for introducing them to the relevant results in \cite{deletingPatternMultiD}, which inspired this work, as well as Zohar Yakhini fruitful discussions about relevant applications.
    
\renewcommand*{\refname}{Bibliography}
\bibliographystyle{plain}
\bibliography{dna-design-ref}


\appendix
\newpage
 		\section{Eliminating unwanted patterns over binary alphabet}\label{EliminatingOverBin}
	 	While the elimination problem for non-binary alphabet sequences is addressed by Algorithm \ref{algBasicAlgForEliminating} (see Section \ref{ConnectionBetweenElimAndHitting}), the elimination problem for binary alphabet sequences cannot be solved by the same algorithm. Recall that Algorithm \ref{algBasicAlgForEliminating} uses the positions in the hitting set as the positions in the eliminating set. However, over the binary alphabet, flipping a position in the hitting set can create a new $P$-match, as shown in the example in Figure \ref{fig:EliminatePatternExample}. While alphabets representing molecular data (e.g. DNA) are not binary and do not have this problem, for the sake of theoretical completeness, we devote this section to present a variant of the algorithm for binary alphabet sequences. Our main objective in this section is to figure out a way to modify the minimal hitting set such that:
	 	\begin{enumerate}
	 		\item It remains a hitting set
	 		\item Its size does not increase
	 		\item Flipping bits in the specified positions does not create new $P$-matches.
	 	\end{enumerate}
	 	For this purpose, we distinguish between overlapping matches and non-overlapping matches. Our solution is largely based on the following claim, which is a corollary of Lemma \ref{EliminatingDnaClaim} that also applies to binary alphabets (unlike Claim \ref{corollaryHittingEqualEliminating}).
	 	\begin{claim}\label{Lemma18Claim}
	 		If a position $j$ in $S$ belongs to two or more $P$-matches, then flipping the bit in this position eliminates all $P$-matches that overlap position $j$ and no new $P$-match is created (Lemma $18$ of \cite{deletingPatternMultiD}).
	 	\end{claim}
	 	\begin{proof}
	 		Any bit flipped within a given $P$-match eliminates that $P$-match, so we are left to show that no new $P$-match is created when flipping a bit that belongs to two or more $P$-matches. Let $i_1$ and $i_2$ denote the two starting positions of two overlapping $P$-matches, and assume that flipping the bit in position $j$ creates a new P-match starting at position $i_3$. Consider the sequence $S'$ created from $S$ by flipping position $j$. $S'$ has a $P$-match starting in index $i_3$ and no $P$-matches starting at positions $i_1, i_2$. Flipping $j$ in $S'$ creates two new $P$-matches, starting at positions $i_1$ and $i_2$. 
	 		Since this contradicts Lemma \ref{EliminatingDnaClaim}, we reach a contraindication to our initial assumption that flipping bit $j$ creates a new $P$-match.
	 	\end{proof}

	 	Consider the minimal hitting set, $H$, returned by Algorithm \ref{algEliminatingIntervals}. Claim \ref{Lemma18Claim} implies that if $i\in H$ belongs to more than one $P$-match, then $S_i$ can be flipped without creating a new $P$-match. We are left to handle the indices that belongs to a single $P$-match. Recall that Algorithm \ref{algEliminatingIntervals} always selected the right-most index in a $P$-match. So, a position in $H$ does not belong to an overlap if it belongs to an isolated $P$-match or if it belongs to a $P$-match that only has left overlaps. In the second case, we simply replace index $i$ with index $i-k+1$. Index $i$ covered only one $P$-match and is the right-most index of that $P$-match, then $i-k+1$ is the left-most index of the same $P$-match. Since this $P$-match has a left overlap, then $i-k+1$ belongs to more than one $P$-match. Therefore, replacing index $i$ with index $i-k+1$ maintains the hitting set, and since index $i-k+1$ belongs to an overlap, Claim \ref{Lemma18Claim} guarantees that we can flip it without creating new $P$-matches.
	 	
	 	We are left to deal with isolated $P$-matches. For this purpose, we utilize the observation made in \cite{deletingPatternMultiD} (Theorem $9$), stating that for all but four degenerate patterns $01^{k-1}, 10^{k-1}, 0^{k-1}1, 1^{k-1}0$, there is a position in each $P$-match that can be flipped without creating a new $P$-match. For nearly all non-degenerate patterns the offset of this position relative to the starting index of the $P$-match depends only on $P$ and is constant across $P$-matches. We describe here how to compute the offset for a given (non-degenerate) pattern. The offset is computed by considering the first bit in $P$ ($b \in \{0,1\}$) and examining the longest substring in $P$ that does not contain $b$ ($\bar{b}$-streak); let $t$ denote the length of this $\bar{b}$-streak.  
	 	\begin{itemize}
	 		\item Case a: $P \in \{0^k, 1^k\}$: the offset is set to $1$.
	 		\item Case b: There is a $\bar{b}$-streak that ends in position $j < k$ in $P$ (not a suffix): the offset is set to $j+1$.
	 		\item Case c: The only $\bar{b}$-streak is a suffix of $P$ but $P \ne b^{k-t}\bar{b}^t$: the offset is set to be the index of the left-most $\bar{b}$ in $P$.
	 		\item Case d: $P=b^{k-t}\bar{b}^t$, where $1 < t < k-1$: if the $P$-match is not in the beginning of $S$ and the bit before the $P$-match is $\bar{b}$, then the offset is set to $2$, otherwise, it is set to $1$. \\
	 	\end{itemize}
 	 	
	 	For an isolated $P$-match starting in position $i$, we compute the relevant offset and add the position $i+offset(i)-1$ to the hitting set (instead of the rightmost position selected by Algorithm \ref{algBasicAlgForEliminating}). Note that for most $P$-matches in $S$, $offset(i)$ does not depend on the location of the specific $P$-match, and only in case d $offset(i)$ can be either $1$ or $2$ depending on $S_{i-1}$. A case-by-case analysis shows that flipping the bit in that position does not generate a new $P$-match (see Theorem $1$ in \cite{deletingPatternMultiD})
	 	
	 	We presented two simple modifications to the minimal hitting set 
	 	returned by Algorithm \ref{algEliminatingIntervals} that produce a minimal hitting set that is also an eliminating set. The modification requires:
	 	\begin{enumerate}
	 		\item Identifying the isolated $P$-matches
	 		\item Selecting a position to substitute in each isolated $P$-match
	 		\item Identifying $P$-matches that have only left overlaps
	 	\end{enumerate}
	 	The added complexity of these steps is $O(k+|H|)$. 

	 	There are four degenerate patterns that are not handled in the analysis done in \cite{deletingPatternMultiD} and in the modified algorithm we described above: $\{01^{k-1}$ , $10^{k-1}$, $0^{k-1}1$, $1^{k-1}0\}$. With degenerate patterns there are cases in which every bit we flip in a $P$-match creates a new $P$-match. Consider, for example, the unwanted pattern $P=0001$ and the sequence $S=0000001001$. The sequence $S$ has a single $P$-match in positions $4-7$. If we flip a bit in position $j\in{4,5,6}$ (from $0$ to $1$), then we create a new $P$-match ending in position $j$. On the other hand, if we flip the bit in position $j=7$ (from $1$ to $0$), we create a new $P$-match ending in position $10$. Indeed, to eliminate $P$ from $S$ we need to flip two bits (e.g. positions $7$ and $10$). This example demonstrates that eliminating degenerate patterns may require more substitutions than the size of the smallest hitting set. Therefore, Algorithm \ref{algBasicAlgForEliminating} from Section \ref{ConnectionBetweenElimAndHitting} is not appropriate in this case and a different algorithmic approach is needed. On the other hand, an algorithm for eliminating degenerate patterns may exploit their special attributes, such as the fact that degenerate patterns cannot have overlapping matches.

\end{document}